\documentclass{article}
\usepackage{epsf}
\begin{document}
\newcommand{\nc}{\newcommand}
\nc{\bea}{\begin{eqnarray}} \nc{\eea}{\end{eqnarray}}
\nc{\beq}{\begin{equation}} \nc{\eeq}{\end{equation}}
\nc{\rr}{\rightarrow}
\nc{\IH}{{}^1{\rm H}}
\nc{\D}{{\rm D}}
\nc{\EH}{{}^3{\rm H}}
\nc{\EHe}{{}^3{\rm He}}
\nc{\UHe}{{}^4{\rm He}}
\nc{\GLi}{{}^6{\rm Li}}
\nc{\ZLi}{{}^7{\rm Li}}
\nc{\ZBe}{{}^7{\rm Be}}
\nc{\DeH}{{\rm D}/{\rm H}}
\nc{\EHeH}{^3{\rm He}/{\rm H}}
\nc{\ZLiH}{{}^7{\rm Li}/{\rm H}}
\nc{\lgZLiH}{\log_{10}({}^7{\rm Li}/{\rm H})}
\nc{\et}{\eta_{10}}

\title{Big Bang Nucleosynthesis Calculation\footnote{Talk given at
"Matter in the Universe", ISSI Bern}}
\author{Hannu Kurki-Suonio \\
Helsinki Institute of Physics and Department of Physics \\
 University of
Helsinki, P.O.Box 64, FIN-00014 Helsinki, Finland}
\date{October 10, 2001}
\maketitle

\begin{abstract}
I review standard big bang nucleosynthesis and some versions of nonstandard
BBN. The abundances of the primordial isotopes D, He-3, and Li-7 produced in
standard BBN can be calculated as a function of the baryon density
with an accuracy of about 10\%.  For He-4 the accuracy is better than 1\%.
The calculated abundances agree fairly well with observations, but the baryon
density of the universe cannot be determined with high precision.
Possibilities for nonstandard BBN include inhomogeneous and antimatter BBN
and nonzero neutrino chemical potentials.
\end{abstract}

\section{Introduction}

Big bang nucleosynthesis (BBN) is among the main observational evidence
for big bang.  The discovery of the cosmic microwave background (CMB)
provided us with the temperature scale of the early universe, and allowed
the calculation of the primordial nuclear abundances produced in the big bang.
The four light isotopes, $\D$, $\EHe$, $\UHe$, and $\ZLi$ are mainly produced
in the big bang, and the calculated abundances agree fairly well with
astronomical observations.

Standard big bang
nucleosynthesis (SBBN) has a single free parameter, the
baryon-to-photon ratio,
\begin{equation}
   \eta \equiv \frac{n_b}{n_\gamma} = 10^{-10}\ldots10^{-9},
\end{equation}
which is related to the present baryonic contribution to the
critical density  $\Omega_b$ via the Hubble constant
$H_0 \equiv h 100$~km/s/Mpc by
\begin{equation}
  \et \equiv 10^{10}\eta = 274\Omega_b h^2.
\end{equation}
For decades, BBN has provided the best determination of the amount of
baryonic matter in the universe.
The agreement with observations is obtained in the range
$\et = 1.5\ldots6$.  Despite optimistic claims from time to time, BBN
has not really progressed towards a much more precise determination
of $\eta$.  Observers claim higher precision from determinations of
primordial abundances of single isotopes, but disagree with each other
or, within the context of SBBN, with primordial abundances of other
isotopes.  Difficult questions about systematic errors in observations
and chemical evolution relating the present abundances to primordial abundances
have prevented further progress.

During the past year, a competing method for estimating the amount of baryonic
matter has appeared.  In the angular power spectrum of the anisotropy of CMB,
the relative heights of the even and odd acoustic peaks are sensitive
to the baryon-to-photon ratio.  The first preliminary estimates from
the Boomerang \cite{Boom00} and Maxima-1 \cite{Maxima00} experiments
appeared to be in conflict with
BBN, giving a higher baryon density, $\Omega_b h^2 \sim 0.03$,
or $\et \sim 8$  \cite{BooMax00}.
The Boomerang collaboration has since revised
their estimate downward, to $\Omega_b h^2 =
0.022^{+0.004}_{-0.003}$ \cite{Boom01}, which agrees with SBBN,
but the Maxima-1 estimate
has been revised upward to $\Omega_b h^2 =
0.0325\pm0.0125$ ($95\%$ c.l.) \cite{Maxima01}.
With the coming satellite experiments
CMB may surpass BBN as the method for estimating $\eta$.  BBN will then
become a
tool for understanding the astrophysics of chemical evolution,
by telling us the primordial abundances.

While SBBN is simple and natural, and is at present in reasonable
agreement with observations, there is interest in studying nonstandard
variants of BBN.  For one thing, BBN is a sensitive probe of the physics of
the early universe.  If we change something about our assumptions
regarding the conditions in the early universe, or the physics
relevant at that time, we are likely to change the primordial abundances
and ruin the agreement with observations. Thus for many things BBN
provides the strongest constraint.

On the other hand, from time to time there have been suggestions for disagreement
between the estimated primordial abundances of the different
isotopes, and/or other ways of estimating $\eta$.  If such disagreements
persist, nonstandard BBN (NSBBN) may be the solution.

I shall begin with a review of the physics of SBBN, and then discuss
a few NSBBN scenarios.

\section{Physics of Big Bang Nucleosynthesis}

In the early universe the temperature is falling as the universe
expands.  The time scale depends on the number of particle species which
are relativistic at that time.  In the standard case these are electrons,
positrons, photons, and 3 species of neutrinos.

There is lots of radiation and very little matter. The amount of baryonic
matter is not known exactly, and is given by $\eta$, the only free
parameter in SBBN.

Weak reactions are converting neutrons into protons.  At first the reaction
rate is high enough to maintain the equilibrium neutron-to-proton
ratio
\begin{equation}
  \biggl(\frac{n}{p}\biggr)_{\rm eq} = e^{-(m_n-m_p)/T-\xi_e},
  \label{npeq}
\end{equation}
which is falling with temperature.  Here $\xi_e \equiv \mu_{\nu_e}/T$ is
the electron neutrino degeneracy parameter ($\mu_{\nu_e}$ is the electron
neutrino chemical potential).  For $\xi \ll 1$, we have roughly
\beq
   \xi \sim \frac{n_\nu - n_{\bar{\nu}}}{n_\gamma}.
\eeq
In SBBN we assume homogeneous conditions with $\xi_e \sim 0$.

At a temperature of about $T \sim 0.8$~MeV the neutrinos decouple and after
that the neutron abundance evolves via free neutron decay
\begin{equation}
   n \rightarrow p + e^- + \bar{\nu}_e.
\end{equation}

Nuclear reactions begin by neutrons and protons producing deuterium.  This
reaction keeps the deuterium abundance close to its equilibrium value.
Because of the large amount of photons in the background radiation and the
low binding energy, $B_d = 2.22$~MeV, of deuterium, the deuterium abundance does not
become large until the temperature has fallen to about $70$~keV.
Only then can the nuclear reactions proceed beyond deuterium.

As the temperature falls further the Coulomb barrier shuts down the nuclear
reactions.  Because of the short time and bottlenecks due to the lack of
stable nuclei at masses $A = 5$ and $A = 8$, the reactions do not
proceed beyond $A = 7$.

There are 10 important reactions which take nucleosynthesis beyond deuterium:
$$\begin{array}{llll}
   d(p,\gamma)\EHe & d(d,p)t & d(d,n)\EHe & \EHe(n,p)t \\
   t(d,n)\UHe & \EHe(d,p)\UHe \quad & \EHe(\alpha,\gamma)\ZBe \quad
   & t(\alpha,\gamma)\ZLi \\
   \ZBe(n,p)\ZLi \quad & \ZLi(p,\alpha)\UHe \nonumber
\end{array}$$
Afterwards tritium $\beta$-decays into $\EHe$ and $\ZBe$ becomes $\ZLi$ by
electron capture.

Since $\UHe$ has the highest binding energy per nucleon
(for $A \leq 7$), almost all neutrons end up incorporated in $\UHe$.
Thus the primordial $\UHe$ abundance $Y_p$ is determined by the neutron
fraction at nucleosynthesis time.  This in turn is determined by the
competition between the weak reaction rates and the rate at which the
temperature falls due to the expansion of the universe.  A higher baryon
density causes nucleosynthesis to take place at a higher
temperature, when there are more neutrons left.  Thus $Y_p$ increases with
$\eta$.

A higher $\eta$ means more efficient
nuclear burning to $\UHe$, leaving less of the
``impurities", $\D$ and $\EHe$.  There is also less directly produced $\ZLi$,
but a higher density allows more $\ZBe$ to be produced, so that the final
primordial $\ZLi$ abundance has a more complicated dependence on $\eta$
(see figure 1).

\section{Accuracy of Nucleosynthesis Yields}

The accuracy of the SBBN calculation of produced primordial abundances depends
on how accurately the various thermonuclear reaction rates are known.
These rates cannot be calculated from first principles at present, and are thus
based on laboratory measurements.  BBN codes make use of compilations
of these rates \cite{FCZ67,FCZ75,HFCZ83,CFHZ85,CF88}.
The Caughlan\&Fowler \cite{CF88} 1988 compilation was updated on some key
BBN rates by Smith et al.~\cite{SKM93}.
The latest compilation is by the NACRE collaboration \cite{NACRE99}.

For BBN, the most significant difference between the new NACRE rates and
the older rates is that deuterium production is now higher.
With the NACRE rates, the O'Meara et al.~\cite{OMeara} result $\DeH =
3.0\pm0.4\times10^{-5}$ corresponds to $\et = 5.9\pm0.5$ instead of
$\et = 5.6\pm0.5$ obtained with the older rates.

Based on how accurately the reaction rates are known, Cyburt et al.~\cite{CFO01}
have estimated the accuracy of SBBN yields in the range $\et = 1\ldots10$
to be better than $0.3$~\% for $Y_p$, $3$--$7$~\% for D, $3$--$10$~\%
for $\EHe$, and $12$--$19$~\% for $\ZLi$.

\section{Observations}

\begin{figure}
\centerline{
\epsfysize=14.0cm
\epsffile{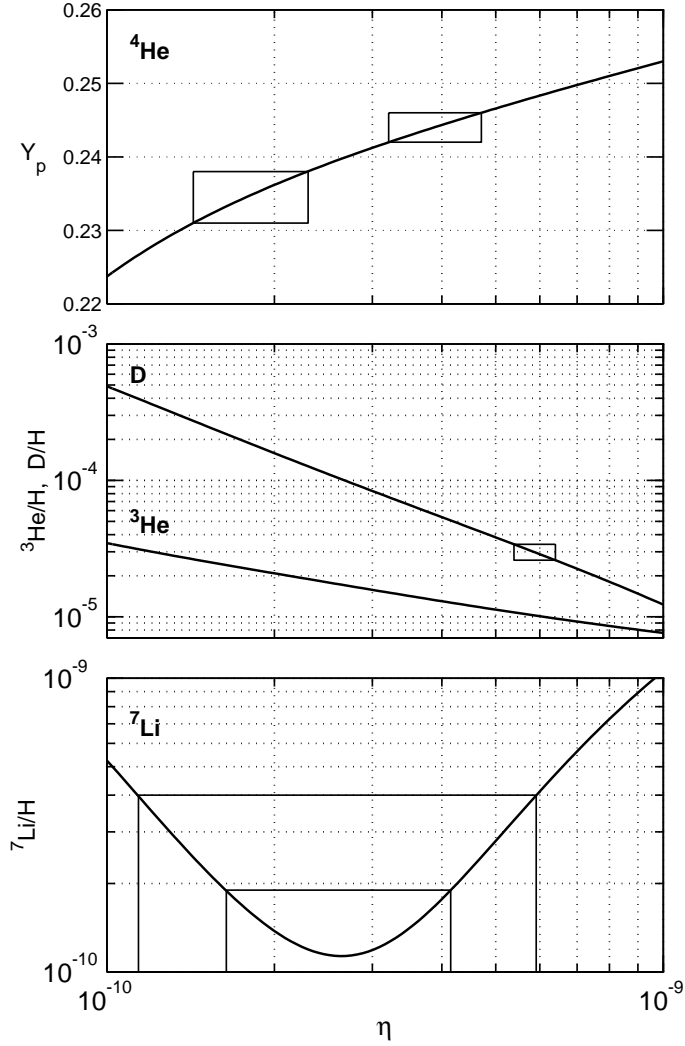}
}
\caption{The SBBN yields of the light isotopes as a function of
the baryon-to-photon ratio $\eta$.  The rectangles corresponds to the
various observational estimates for the primordial abundances mentioned in the
text: $\UHe$ is from Olive et al.~\protect\cite{OSS97} and Izotov\&Thuan
\protect\cite{IZ98},
$\D$ is from O'Meara et al.~\protect\cite{OMeara} and $\ZLi$ is from
Ryan et al.~\protect\cite{Ryan00}.
For $\ZLi$ a conservative upper limit
$\lgZLiH \leq -9.4$ is also shown.  Figure from \protect\cite{Sihvola01a}.}
\end{figure}

I shall leave the more detailed discussion of the observations relating to
primordial abundances to other participants of this workshop, and just list
various recent results in the literature.

For $\UHe$, Olive et al.~\cite{OSS97} combined results of different observers
and estimated a primordial abundance $Y_p = 0.234\pm0.003$.  Izotov\&Thuan \cite{IZ98}
used their own large set of observations to arrive at $Y_p = 0.244\pm0.002$.
The errors are $1$-$\sigma$ statistical errors.  There has been much recent
discussion of possible systematic errors.
Peimbert et al.~\cite{Peimbert00} have recently measured the
$\UHe$ abundance in the SMC, whose proximity allows a better control
of systematic errors. They obtained $Y = 0.2405\pm0.0018$ for the SMC,
and from this they estimate $Y_p = 0.2345\pm0.0026$ for the primordial abundance.
Thuan\&Izotov \cite{TIhere} have now refined their estimate to $Y_p= 0.2443\pm0.0015$.
After discussing various systematic effects, they conclude that because of
systematic errors this could be an underestimate by $\sim$ 2--4\%.

The observed deuterium abundance in the local interstellar medium
is $\DeH = 1.50\pm0.10\times10^{-5}$ \cite{Linsky98}, but there is
evidence for local variations \cite{VidalM01}.  For example,
Sonneborn et al.~\cite{Sonneborn00} have recently reported
$\DeH = 2.18^{+0.36}_{-0.31}\times10^{-5}$ along one line of sight.
Since deuterium is destroyed
in stellar processes, its primordial abundance should be greater than the
local abundance.

The most promising method of obtaining the primordial deuterium abundance
is the measurement of $\DeH$ from Lyman-series absorption by high-redshift
clouds.  Based on three such measurements and upper limits from others,
O'Meara et al.\cite{OMeara} obtain $\DeH = 3.0\pm0.4\times10^{-5}$.
Measurements on
one such system suggest a higher deuterium abundance, possibly larger than
$10^{-4}$ \cite{Webb,Tytler99}.
Recently, the deuterium in two more such systems has been observed
\cite{DOdorico,Levshakov,Pettini},
supporting the low $\DeH$ of O'Meara et al.~\cite{OMeara}.

The estimates for the primordial $\ZLi$ abundance are based on the rather
uniform abundances in population II halo stars.  Bonifacio\&Molaro \cite{BM97}
obtained $\lgZLiH = -9.80\pm0.012\pm0.05$ and Ryan et al.~\cite{Ryan99} obtained
$\lgZLiH = -9.88$ for the mean abundance.  This measured abundance
is close to the minimum $\ZLi$ from SBBN (see figure 1).  It may be
depleted somewhat from the primordial abundance by stellar processing.
Pinsonneault et al.~\cite{PWSN99} estimate $\lgZLiH = -9.65\ldots-9.25$ for the primordial
abundance.  The observed $\ZLi$
may also include a galactic contribution, so that the primordial abundance
could be even lower.  For the primordial abundance Ryan et al.~\cite{Ryan00}
estimate $\lgZLiH = -10.04$\ldots$-9.72$
and Suzuki et al.~\cite{Suzuki00} estimate $\lgZLiH =
-9.97\ldots-9.77$.

Comparing these estimates for primordial abundances to the SBBN yields (figure
1) we see that there is some tension between $\DeH$, which favors a higher
baryon density, and $Y_p$ and $\ZLiH$, which favor a lower baryon density.  The
best estimate from CMB now agrees with the high $\eta$ from $\DeH$.
Thuan\&Izotov \cite{TIhere} now conclude that their value for $Y_p$ is in good agreement
with this when one allows for the systematic error.  If one accepts the
Thuan\&Izotov \cite{TIhere} result for $Y_p$ the remaining disagreement is with the
$\ZLi$.  Possibly $\ZLi$ processing is not yet understood well enough, and the
tighter limits above are too stringent.

\section{Nonstandard BBN}

There are many proposed possible modifications to SBBN.  I shall go over
just a few of these NSBBN scenarios.

A higher energy density at a given temperature would lead to faster expansion
in the early universe.  This would be caused by the presence of additional
relativistic particle species, e.g., additional neutrino species.
The shorter time scale would mean that there are more neutrons left at
nucleosynthesis time, and thus more $\UHe$ is produced.  This is the most
important effect.  Since a higher $Y_p$ tends to lead to worse agreement with
observations, BBN sets an upper limit to the speed-up possible.  This upper
limit is usually parameterized in terms of the effective number of (light)
neutrino species $N_\nu$.  Different authors get different upper limits
depending on the observational constraints chosen.  Two recent results are
$N_\nu < 3.2$ by Burles et al.~\cite{Burles99} and $N_\nu \leq 4$ by
Lisi et al.~\cite{Lisi99}.

The effect on other isotopes is roughly that their abundance curves vs. the
baryon density are shifted towards higher $\eta$.  The higher density
compensates for the shorter time available.  This is a smaller effect than
the effect on $\UHe$.

A large neutrino degeneracy, $\xi \neq 0$, would increase the energy density
in neutrinos at a given temperature, also leading to a speed-up of the
expansion rate.  If the degeneracy is in electron neutrinos, a much larger
effect is that $\xi_e\neq 0$ shifts the balance of weak reactions (see
eq.~\ref{npeq}).  If $\xi_e > 0$, so that there are more electron neutrinos than
electron antineutrinos, we get less neutrons and thus a lower $Y_p$.
Since this has a large effect on $Y_p$ but a small effect on the other
isotopes, we can use $\xi_e$ to ``dial in'' a desired value of $Y_p$,
as $\xi_e$ is otherwise unobservable.

One can combine the above scenarios by having both a significantly
faster expansion rate and $\xi_e > 0$, so that the effects on $Y_p$ cancel each
other to keep it in the observational range, while the effect of the
speed-up on the other isotopes remains \cite{KS92}.
If one wants to use neutrino degeneracy only to achieve this speed-up, this
scenario requires $|\xi_\mu| \gg  \xi_e $ or $|\xi_\tau| \gg  \xi_e $.
This way larger $\eta$ are allowed than in SBBN.  However, since one is relying
on a small effect, the large effects having cancelled each other, this scenario
requires a very large speed-up for a significant effect.  A large $N_\nu$ leads
to other cosmological effects which put limits to this scenario.  In
particular, the recent CMB anisotropy results place an upper bound to $N_\nu$
 \cite{Hannestad00,Hannestad01}.
When the CMB constraints are combined with this
BBN scenario \cite{Esposito00,Kneller01,LL01,Hansen01}
the conclusion is that the
upper limit to $\et$ can be raised
from the SBBN $\et < 6$ to $\et \leq 7\ldots 8$ only \cite{Kneller01}.

\section{Inhomogeneous BBN}

\begin{figure}
\centerline{
\epsfysize=7.0cm
\epsffile{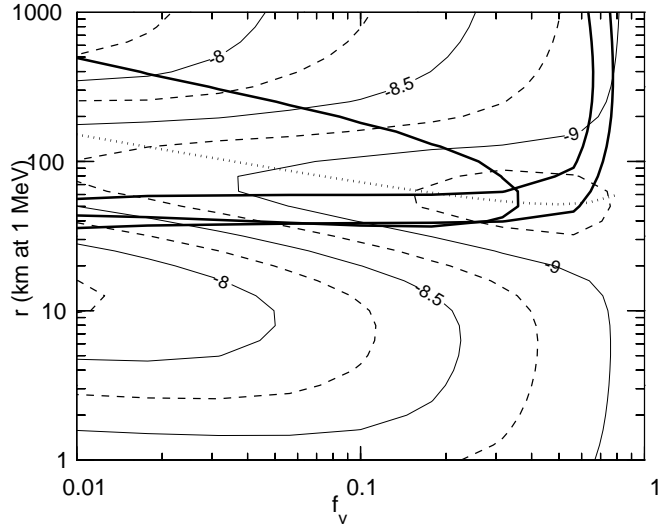}
}
\caption{Results from IBBN for $\Omega_b h^2 = 0.030$ ($\et = 8.22$).
This is the density suggested by the preliminary Boomerang and Maxima-1
results.
$r$ is the distance scale giving the separation between centers of high-
and low-density regions, and $f_v$ is the volume fraction of the high-density
regions.  The distance scale is given in comoving units at $T = 1$~MeV.
($1$~km at $1$~MeV corresponds to $1.9\times10^{-4}$~pc at present.)
The thick lines show the observational constraints $Y_p \leq 0.248$
(to the left of the curve) and $\DeH = 2.9$--$4.0\times10^{-5}$.  IBBN can
thus bring this high $\eta$ into agreement with $\UHe$ and $\D$ observations.
Thin lines are contours of $\lgZLiH$, showing that $\ZLi$ remains
problematically high.  This figure is for a ``spherical shell'' geometry
which is more successful in  reducing the $\ZLi$ yield than other
geometries tried.  The dotted curve shows where figure 3 cuts through
this figure. Figure from \protect\cite{ibbnboom}.}
\end{figure}

\begin{figure}
\centerline{
\epsfysize=7.0cm
\epsffile{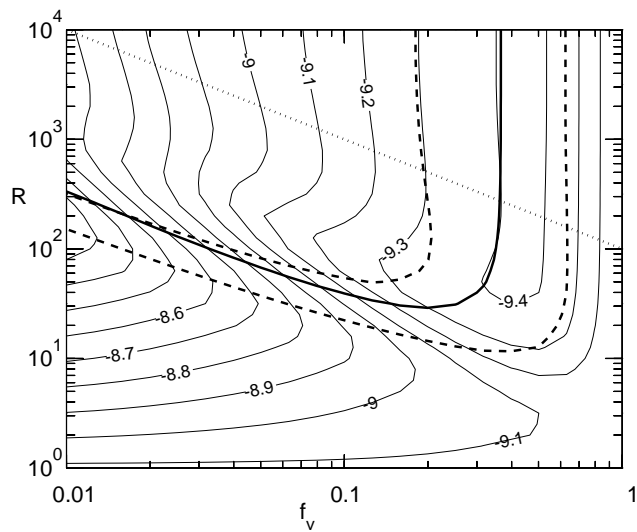}
}
\caption{Like figure 2, but now we show the dependence on the
third IBBN parameter, the density contrast $R$ between
the high- and low-density regions.  The dotted line shows where
figure 2 cuts through this one.  The solid thick line is the constraint
$Y_p \leq 0.248$ (above the line) and the dashed thick lines give the
$\DeH$ constraint.  The thin lines are the $\lgZLiH$ contours.  This
model can just reach below our ``conservative upper limit''
$\lgZLiH \leq -9.4$. Figure from \protect\cite{ibbnboom}.}
\end{figure}

Another way to modify SBBN is to relax the homogeneity assumption, and allow
$\eta$ to be a function of position at nucleosynthesis time.
In the usual kind of inhomogeneous BBN (IBBN), the baryon density is
positive, $\eta > 0$, everywhere, but one can also consider the possibility of
having antimatter regions, where $\eta < 0$.  This
latter case I call antimatter BBN
(ABBN), and I discuss it in the next section.

The crucial question in IBBN is the distance scale $r$
of the inhomogeneity.  If the mechanism causing the inhomogeneity is connected
with inflation, inhomogeneity at any scale could be produced.  The observed
isotropy of the CMB, however, requires $\eta$ to be highly homogeneous
at distance scales larger than about 10 Mpc.  Thus IBBN cannot be used
to explain, e.g., different deuterium abundances at different
high-redshift Lyman-absorbers.

In the usual IBBN scenarios one assumes a much smaller distance scale of the
inhomogeneity, so that while $\eta$ is inhomogeneous at nucleosynthesis time,
the matter from different regions gets mixed later, resulting in a homogeneous
baryon density with homogeneous abundances, which however are different
from the SBBN abundances for the same $\eta$.

In the simplest version of IBBN, each small region undergoes SBBN with
its own baryon-to-photon ratio, and one gets the final abundances
by averaging over the SBBN abundances over the distribution in $\eta$.
Leonard\&Scherrer \cite{LS96} have shown that this way one can relax, or even remove, the
observational lower limit to $\eta$, but the upper limit to $\eta$
remains unchanged.

If the mechanism producing the inhomogeneity was not connected with inflation,
causality requires the distance scale to be smaller than the horizon at that
time.  A favorite candidate for producing the inhomogeneity has been the QCD
phase transition at $T \sim 150$~MeV, when the horizon was about $1$~pc
(comoving).  The order of the phase transition is not known, but if it was of
first order, it would have proceeded through nucleation of bubbles of hadronic
matter in the ambient quark-gluon plasma.  Near the end of the transition
there would have been shrinking droplets of quark-gluon plasma, where much of
the baryon number would have been concentrated due to the difficulty of baryon
number crossing the phase boundary.  This way very high density contrasts could
be produced.  The relevant distance scale is close to the neutron diffusion
scale at nucleosynthesis time.  An inhomogeneity of this scale would produce
a strongly
inhomogeneous neutron-to-proton ratio due to neutron diffusion out of the
high-density regions, leading to a drastic change in BBN \cite{AHS87,
AFM87,MF88}.  In the first papers the effect
was overestimated; for more accurate calculations see, e.g.,
\cite{K88,Mathews90,JFM94}. A more extensive list of references is
given in the review articles by Reeves \cite{Reeves91} and Malaney\&Mathews \cite{MM93}
and in, e.g., \cite{KKS99}.  For a range of distance scales one can get
less $\UHe$ and more $\D$ than in SBBN, favoring a higher $\eta$.
One can also get some reduction in the $\ZLi$ yield, but not much.  Because
of the large number of parameters in the scenario, quantitative results
are not easily summarized.  The IBBN scenario is contrasted with
recent observations in \cite{KKS99} and \cite{ibbnboom}.

The properties of the QCD phase transition can be estimated by lattice QCD
calculations.  While a distance scale interesting for IBBN can not be ruled
out, unfortunately a too small distance scale appears more likely.

\section{BBN with Antimatter Regions}

We do not know the origin of the baryoasymmetry, $\eta > 0$, of the universe.
Theories of baryogenesis try to explain this excess of matter over antimatter
and to predict (postdict) the observed value of $\eta$.
While the simplest models usually lead to a homogeneous $\eta$, in many
baryogenesis models the baryon density may come out inhomogeneous, and
in some regions the asymmetry may have the opposite sign, so that after local
annihilation we end up with regions of antimatter \cite{Dolgov}.

\begin{figure}
\centerline{
\epsfysize=5.0cm
\epsffile{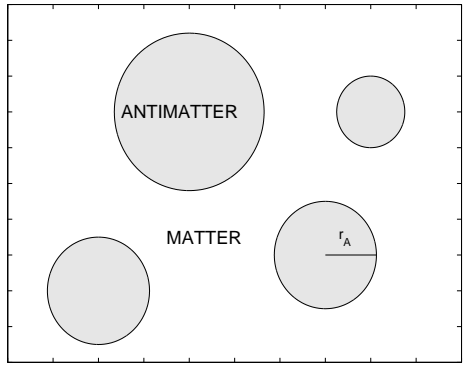}
}
\caption{Antimatter regions. Figure from \protect\cite{Sihvola01a}.}
\end{figure}

Annihilation will then proceed at the matter-antimatter boundary.
The smaller the antimatter regions the sooner they are completely annihilated.
From the spectrum of the CMB we know that there was no major annihilation
going on close to recombination time, and the observed cosmic diffuse
gamma ray radiation puts tight limits on annihilation after recombination.
This leaves
us three possibilities not in contradiction with observations:
1) the antimatter regions annihilated well before recombination time,
2) the amount of antimatter was much less (a factor of about $10^{-6}$
or less) than the amount of matter, or
3) the antimatter regions are separated by large distances, comparable to the
present horizon, or larger.

The first possibility leads to an interesting NSBBN scenario, antimatter
BBN (ABBN) \cite{RJ98,RJ98,RJ01,abbnl,abbnl,abbn,Sihvola01}.  We consider
antimatter regions which are larger than $10^{-5}$~pc but smaller than
$100$~pc (comoving).  These annihilate after $T = 1$~MeV, but before recombination,
and the annihilation process can significantly affect nucleosynthesis, or
modify the abundances afterwards.  Since in the end we have complete
annihilation of the antimatter, we need to assume an excess of matter
over antimatter, so that the antimatter/matter ratio $R < 1$.

Antimatter regions smaller than $10^{-2.5}$~pc annihilate before
nucleosynthesis, when the temperature of the universe is $ T =
1$~MeV--$70$~keV.  The mixing of matter and antimatter is due to (anti)neutron
diffusion, and therefore the annihilation is preferably on neutrons.
This reduces the $\UHe$ production.  The observational lower limit to $Y_p$
therefore provides an upper limit to the amount of antimatter $R$.

\begin{figure}
\centerline{
\epsfysize=7.0cm
\epsffile{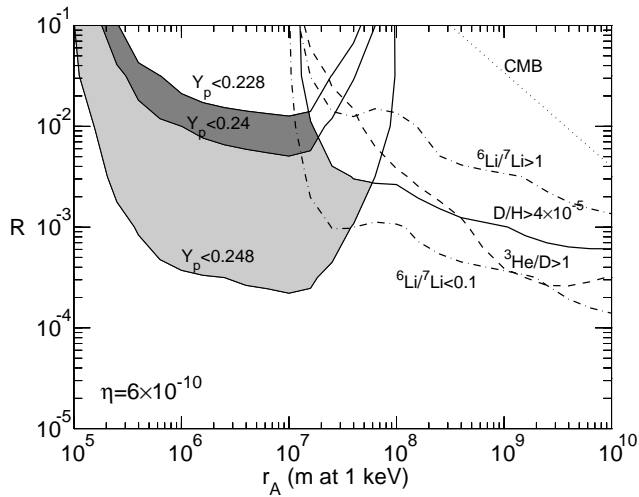}
}
\caption{Observational constraints for ABBN with $\et = 6$.
$R$ is the antimatter/matter ratio,
and $r_A$ is the radius of the antimatter region, given in comoving units
at $T = 1$~keV.
($1$~m at $1$~keV is $4.26\times10^6$~m $= 1.38\times10^{-10}$~pc today.)
The shaded
region is the region allowed by the constraints $Y_p = 0.228$--$0.248$
and $\DeH = 2.2$--$4.0\times10^{-5}$.  ABBN is thus able to remove the
tension between $\D$ and $\UHe$ observations.  This region is also allowed
by other constraints: the spectral distortion of the CMB and upper limits
to $\EHe/\D$; the excluded regions are above and to the right
of these curves.  Because $\GLi$ is very fragile, it is difficult to
make definitive conclusions about its primordial abundance from observations.
Therefore we just show two contours, $\GLi/\ZLi = 0.1$ and $\GLi/\ZLi = 1$.
The $\ZLi$ yield is fairly constant, close to the SBBN
value $\ZLiH = 2.8\times10^{-10}$, over the
whole figure.
Figure from \protect\cite{Sihvola01}.}
\end{figure}

Larger antimatter regions survive until nucleosynthesis, which consumes the
free neutrons.  This stops further annihilation for a while.  There is then
a second burst of annihilation well after nucleosynthesis, when proton
diffusion becomes effective, at $T < 3$~keV.  The annihilation process
then changes the nuclear abundances through several mechanisms.

Gamma rays from annihilation photodisintegrate nuclei.  The major effect
is the production of $\D$ and especially $\EHe$ from
photodisintegration of $\UHe$.

Antiproton annihilation on helium nuclei produces $\EH$, $\EHe$, $\D$,
protons and neutrons.  These nuclear remnants have high energies,
and the energetic $\EH$ lead to nonthermal production of $\GLi$
by the endoergic reaction
$t(\alpha,n)\GLi$.

Thus $\D$, $\EHe$, and $\GLi$ abundances are increased over the SBBN
yields.  There is no big effect on $\ZLi$.

At $T < 1$~keV, the energy released in annihilation does not get fully
thermalized, but results in a distortion of the CMB spectrum.  Since no such
distortion has been observed, CMB places constraints an the amount of
antimatter annihilating below $T \sim 1$~keV, but before recombination.
Near $T \sim 1$~keV the universe is still strongly radiation dominated,
so the energy release in annihilation is small compared to the
energy in the background radiation.  As the temperature falls, the
matter-to-radiation energy ratio increases. Therefore the limits from CMB
to the amount of antimatter
become progressively stronger for larger distance scales, for which
the annihilation occurs later.  For scales smaller than $1$~pc, we get,
however, a stronger limit from $\EHe$ overproduction.

\begin{figure}
\centerline{
\epsfysize=7.0cm
\epsffile{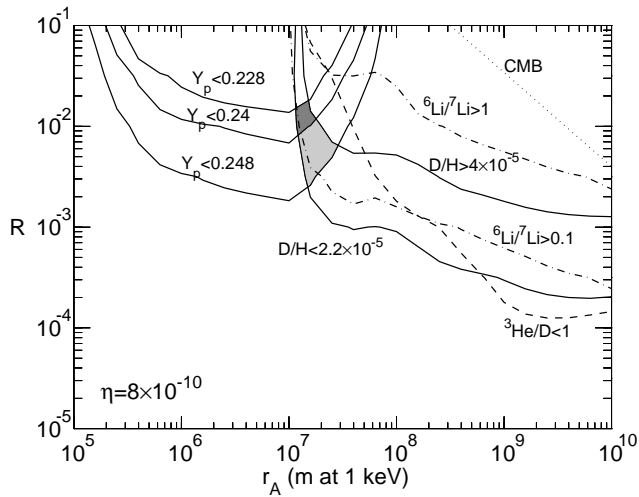}
}
\caption{Same as the previous figure, but with $\et = 8$.  ABBN can accommodate
higher values of $\eta$ than SBBN, except for the $\ZLi$ constraint,
which is essentially the same for ABBN and SBBN.
Figure from \protect\cite{Sihvola01}.}
\end{figure}

Since ABBN reduces $Y_p$ and increases $\DeH$, it allows for a larger
baryon density, at least from those observational constraints.  The
constraint from $\ZLi$, however, remains essentially unchanged.

\section{Conclusions}

Standard BBN is compelling in its simplicity.
While there is controversy among the observers and some apparent discrepancy
between the estimated primordial abundances of the different isotopes
and SBBN, these are probably not serious, and most likely represent
difficulties in making observations and estimating primordial abundances
based on observed ones.  SBBN is thus not in trouble.  Unfortunately,
because of these difficulties, it is not able to pin
down the baryon-to-photon ratio very precisely.  It is somewhere in
the range $\et = 1.5$--$6$, or $\Omega_b h^2 = 0.005$--$0.022$.
The high redshift deuterium measurements point towards the upper end
of this range, $\et \sim 5$--$6$, or $\Omega_b h^2 \sim 0.02$.

The recent estimates from CMB anisotropy, $\Omega_b h^2 =
0.022^{+0.004}_{-0.003}$ from Boom\-erang \cite{Boom01}
and $\Omega_b h^2 = 0.0325\pm0.0125$ ($95\%$ c.l.) from Maxima-1 \cite{Maxima01}
also favor this upper end of the range.  We are eagerly
waiting for more precise
CMB measurements in the coming years.

While standard BBN is in good shape, there is interest in studying
nonstandard BBN: to assess the robustness of SBBN, to constrain
possibilities for nonstandard physics or cosmology, and to be ready to
provide relief if observational discrepancies turn out to be serious
for SBBN.

We discussed here four NSBBN scenarios: 1) electron neutrino degeneracy,
2) electron neutrino degeneracy combined with a speed-up of the expansion rate
due to extra energy density, 3) inhomogeneous BBN, and 4) antimatter BBN.
All these scenarios are able to relieve the tension between the $\D$ and $\UHe$
observations.  The three last ones may also allow a larger baryon density
than SBBN, but with some difficulty:  2) is constrained in that respect
by other cosmological constraints
and 3) and 4) cannot do much for the $\ZLi$ constraint.

I thank Elina Sihvola for permission to reproduce figures from
\cite{Sihvola01,Sihvola01a}.


\begin{thebibliography}{99}

\bibitem{AFM87}
Alcock, C., Fuller, G.M. and Mathews, G.J.: 1987,
{\it Astrophys. J.} {\bf 320}, 439.

\bibitem{NACRE99}
Angulo, C. et al. (NACRE Collaboration): 1999,
{\it Nucl. Phys.} {\bf A656}, 3.

\bibitem{AHS87}
Applegate, J.H., Hogan, C.J. and Scherrer, R.J.: 1987,
{\it Phys. Rev. D} {\bf 35}, 1151.

\bibitem{Maxima00}
Balbi, A. et al.: 2001,
{\it Astrophys. J.} {\bf 545}, L1.

\bibitem{BM97}
Bonifacio, P. and Molaro, P.: 1997,
{\it Mon. Not. R. Astron. Soc.} {\bf 285}, 847.

\bibitem{Burles99}
Burles, S., Nollett, K.M., Truran, J.W. and Turner, M.S.: 1999,
{\it Phys. Rev. Lett.} {\bf 82}, 4176.

\bibitem{CFHZ85}
Caughlan, G.R., Fowler, W.A., Harris, M.J and Zimmerman, B.A.: 1985,
{\it At. Data Nucl. Data Tables} {\bf 32}, 197.

\bibitem{CF88}
Caughlan, G.R. and Fowler, W.A.: 1988,
{\it At. Data Nucl. Data Tables} {\bf 40}, 283.

\bibitem{CFO01}
Cyburt, R.H., Fields, R.D., Olive, K.A.: 2001,
astro-ph/0102179.

\bibitem{DOdorico}
D'Odorico, S., Dessauges-Zavadsky, M. and Molaro, P.: 2001,
astro-ph/0102162.

\bibitem{Dolgov}
Dolgov, A.D.: 1996,
hep-ph/9605280.

\bibitem{Esposito00}
Esposito, S., Mangano, G., Melchiorri, A., Miele, G. and Pisanti, O.: 2001,
{\it Phys. Rev. D} {\bf 63}, 043004.

\bibitem{FCZ67}
Fowler, W.A., Caughlan, G.R. and Zimmerman, B.A.: 1967,
{\it Ann. Rev. Astron. Ap.} {\bf 5}, 525.

\bibitem{FCZ75}
Fowler, W.A., Caughlan, G.R. and Zimmerman, B.A.: 1975,
{\it Ann. Rev. Astron. Ap.} {\bf 13}, 69.

\bibitem{Hannestad00}
Hannestad, S.: 2000,
{\it Phys. Rev. Lett.} {\bf 85}, 4203.

\bibitem{Hannestad01}
Hannestad, S.: 2001,
{\it Phys. Rev. D} {\bf 64}, 083002.

\bibitem{Hansen01}
Hansen, S.H., Mangano, G., Melchiorri, A., Miele, G. and Pisanti, O.: 2001,
astro-ph/0105385.

\bibitem{HFCZ83}
Harris, M.J., Fowler, W.A., Caughlan, G.R. and Zimmerman, B.A.: 1983,
{\it Ann. Rev. Astron. Ap.} {\bf 21}, 165.

\bibitem{IZ98}
Izotov, Y.I. and Thuan, T.X.: 1998,
{\it Astrophys. J.} {\bf 500}, 188.

\bibitem{BooMax00}
Jaffe, A.H. et al.: 2001,
{\it Phys. Rev. Lett.} {\bf 86}, 3475.

\bibitem{JFM94}
Jedamzik, K., Fuller, G.M., and Mathews, G.J.: 1994,
{\it Astrophys. J.} {\bf 423}, 50.

\bibitem{KKS99}
Kainulainen, K., Kurki-Suonio, H. and Sihvola, E.: 1999,
{\it Phys. Rev. D} {\bf 59}, 083505.

\bibitem{KS92}
Kang, H.-S. and Steigman, G.: 1992,
{\it Nucl. Phys. B} {\bf 372}, 494.

\bibitem{Kneller01}
Kneller, J.P., Scherrer, R.J., Steigman, G. and Walker, T.P.: 2001,
astro-ph/0101386.

\bibitem{K88}
Kurki-Suonio, H., Matzner, R.A., Centrella, J.M., Rothman, T. and Wilson, J.: 1988,
{\it Phys. Rev. D} {\bf 38}, 1091.

\bibitem
{abbnl}
Kurki-Suonio, H. and Sihvola, E.: 2000,
{\it Phys. Rev. Lett.} {\bf 84}, 3756.

\bibitem
{abbn}
Kurki-Suonio, H. and Sihvola, E.: 2000a,
{\it Phys. Rev. D} {\bf 62}, 103508.

\bibitem
{ibbnboom}
Kurki-Suonio, H. and Sihvola, E.: 2001,
{\it Phys. Rev. D} {\bf 63}, 083508.

\bibitem{Boom00}
Lange, A.E. et al.: 2001,
{\it Phys. Rev. D} {\bf 63}, 042001.

\bibitem{LS96}
Leonard, R.E. and Scherrer, R.J.: 1996,
{\it Astrophys. J.} {\bf 463}, 420.

\bibitem{LL01}
Lesgourgues, J. and Liddle, A.R.: 2001,
astro-ph/0105361.

\bibitem{Levshakov}
Levshakov, S.A., Dessauges-Zavadsky, M., D'Odorico, S. and Molaro. P.: 2001,
astro-ph/0105529.

\bibitem{Linsky98}
Linsky, J.L.: 1998,
{\it Space Science Rev.} {\bf 84}, 285.

\bibitem{Lisi99}
Lisi, E., Sarkar, S. and Villante, F.L.: 1999,
{\it Phys. Rev. D} {\bf 59}, 123520.

\bibitem{MF88}
Malaney, R.A. and Fowler, W.A.: 1988,
{\it Astrophys. J.} {\bf 333}, 14.

\bibitem{MM93}
Malaney, R.A. and Mathews, G.J.: 1993,
{\it Phys. Rep.} {\bf 229}, 145.

\bibitem{Mathews90}
Mathews, G.J., Meyer, B.S., Alcock, C.R. and Fuller, G.M.: 1990,
{\it Astrophys. J.} {\bf 358}, 36.

\bibitem{Boom01}
Netterfield, C.B. et al.: 2001,
astro-ph/0104460.

\bibitem{OMeara}
O'Meara, J.M., Tytler, D, Kirkman, D., Suzuki, N., Prochaska,
J.X., Lubin, D. and Wolfe, A.M.: 2001,
{\it Astrophys. J.} {\bf 552}, 718.

\bibitem{OSS97}
Olive, K.A., Steigman, G. and Skillman, E.D.: 1997,
{\it Astrophys. J.} {\bf 483}, 788.

\bibitem{Peimbert00}
Peimbert, M., Peimbert, A. and Ruiz, M.T.: 2000,
astro-ph/0003154.

\bibitem{Pettini}
Pettini, M. and Bowen D.V.: 2001,
astro-ph/0104474.

\bibitem{PWSN99}
Pinsonneault, M.H., Walker, T.P., Steigman, G. and Narayanan, V.K.: 1999,
{\it Astrophys. J.} {\bf 527}, 180.

\bibitem{Reeves91}
Reeves, H.: 1991,
{\it Phys. Rep.} {\bf 201}, 335.

\bibitem
{RJ98}
Rehm, J.B. and Jedamzik, K.: 1998,
{\it Phys. Rev. Lett} {\bf 81}, 3307.

\bibitem
{RJ01}
Rehm, J.B. and Jedamzik, K.: 2001,
{\it Phys. Rev. D} {\bf 63}, 043509.

\bibitem{Ryan99}
Ryan, S.G., Norris, J.E. and Beers, T.C.: 1999,
{\it Astrophys. J.} {\bf 523}, 654.

\bibitem{Ryan00}
Ryan, S.G., Beers, T.C., Olive, K.A., Fields, B.D. and Norris, J.E.:
2000,
{\it Astrophys. J.} {\bf 530}, L57.

\bibitem
{Sihvola01}
Sihvola, E.: 2001,
{\it Phys. Rev. D} {\bf 63}, 103001.

\bibitem
{Sihvola01a}
Sihvola, E.: 2001a,
{\it PhD thesis, Univ. of Helsinki}.

\bibitem{SKM93}
Smith, M.S., Kawano, L.H. and Malaney, R.A.: 1993,
{\it Astrophys. J. Suppl. Ser.} {\bf 85}, 219.

\bibitem{Sonneborn00}
Sonneborn, G., Tripp, T.M., Ferlet, R., Jenkins, E.B., Sofia, U.J.,
Vidal-Madjar, A. and Wo\'{z}niak, P.R.: 2000,
{\it Astrophys. J.} {\bf 545}, 277.

\bibitem{Maxima01}
Stompor, R. et al.: 2001,
astro-ph/0105062.

\bibitem{Suzuki00}
Suzuki, T.K., Yoshii, Y. and Beers, T.C.: 2000,
{\it Astrophys. J.} {\bf 540}, 99.

\bibitem{TIhere}
Thuan, T.X. and Izotov, Y.I.: 2001,
this volume.

\bibitem{Tytler99}
Tytler, D., Burles, S., Lu, L., Fan, X.-M., Wolfe, A. and Savage, B.D.: 1999,
{\it Astron. J.} {\bf 117}, 63.

\bibitem{VidalM01}
Vidal-Madjar, A.: 2001,
astro-ph/0103170.

\bibitem{Webb}
Webb, J.K., Carswell, R.F., Lanzetta, K.M., Ferlet, R., Lemoine, M.,
Vidal-Madjar, A. and Bowen, D.V.: 1999,
{\it Nature} {\bf 388}, 250.

\end{thebibliography}
\end{document}